\documentclass[a4paper,11pt]{article}
\pdfoutput=1
\usepackage{jheppub} 
\usepackage{mathtools}
\usepackage{braket}
\newcommand{\Trh}{T_\text{rh}}
\newcommand{\gs}{g_\star}
\newcommand{\gss}{g_{\star s}}
\newcommand{\arh}{a_\text{rh}}
\newcommand{\rp}{\rho_\phi}
\newcommand{\rR}{\rho_R}
\newcommand{\lhs}{\lambda_{hs}}
\newcommand{\ls}{\lambda_s}

\title{Freezing-in Cannibals\\with Low-reheating Temperature}
\author[a]{Nicol\'as Bernal,}
\author[b]{Esau Cervantes,}
\author[a]{Kuldeep Deka,}
\author[b]{Andrzej Hryczuk}
\affiliation[a]{New York University Abu Dhabi\\
PO Box 129188, Saadiyat Island, Abu Dhabi, United Arab Emirates}
\affiliation[b]{National Centre for Nuclear Research\\
Pasteura 7, 02-093 Warsaw, Poland}

\emailAdd{nicolas.bernal@nyu.edu}
\emailAdd{esau.cervantes@ncbj.gov.pl}
\emailAdd{kuldeep.deka@nyu.edu}
\emailAdd{andrzej.hryczuk@ncbj.gov.pl}

\abstract{The freeze-in mechanism provides a compelling framework for dark matter (DM) production, particularly suited to scenarios involving feeble interactions with the Standard Model (SM). In this work, we 
highlight a possible interplay of a non-instantaneous reheating phase and dark sector self-interactions — specifically $2 \to 3$ and $3 \to 2$ cannibalization processes. As an example we study the freeze-in production of a complex scalar DM candidate stabilized by a $\mathbb{Z}_3$ symmetry permitting cubic self-couplings, enabling number-changing interactions that drive internal thermalization and significantly modify the dark sector number density and temperature evolution. We numerically solve the coupled Boltzmann equations for the DM number density and temperature alongside the evolving SM bath, accurately capturing the dynamics of a prolonged reheating epoch. Our analysis reveals a rich and distinctive phenomenology arising from the interplay between the Universe's thermal history, Higgs portal–mediated production, and cannibalistic self-interactions. Compared to scenarios with instantaneous reheating or negligible self-interactions, our framework opens new viable regions in parameter space—particularly for light DM—potentially within reach of future probes.}

\begin{document}
\begin{flushright}
\end{flushright}
\maketitle

\section{Introduction} \label{sec:intro}
The existence of dark matter (DM) is firmly established through a wide range of astrophysical and cosmological observations, yet its underlying nature remains elusive. While the Weakly Interacting Massive Particle (WIMP) paradigm, predicting thermal freeze-out, long dominated theoretical expectations, the lack of any experimental signal (see e.g. Refs.~\cite{Arcadi:2017kky, Roszkowski:2017nbc, Arcadi:2024ukq} for reviews) has prompted a growing interest in alternative production mechanisms, especially those involving feeble interactions with the Standard Model (SM). Among these, the freeze-in mechanism has emerged as a compelling framework in which DM is gradually populated out of equilibrium through rare interactions with the thermal SM bath~\cite{McDonald:2001vt, Choi:2005vq, Kusenko:2006rh, Petraki:2007gq, Hall:2009bx, Bernal:2017kxu}.

However, even within the freeze-in scenario, DM could feature sizable self-interactions that strongly impact its production. The generation of the DM relic abundance could proceed through $N$-to-$N'$ {\it cannibal} number-changing processes, where $N$ DM particles annihilate into $N'$ of them (with $N > N' \geq 2$)~\cite{Carlson:1992fn, Hochberg:2014dra}. The dominant cannibal interactions typically correspond to 3-to-2 annihilations, which can be realized, for example, if the stability of DM is guaranteed by a $\mathbb{Z}_3$ symmetry~\cite{Choi:2015bya, Bernal:2015bla, Bernal:2015lbl, Ko:2014nha, Choi:2017mkk, Chu:2017msm}. Instead, if DM is stabilized by a $\mathbb{Z}_2$ symmetry, 4-to-2 reactions would be those that give rise to the DM relic abundance, while 3-to-2 annihilations are forbidden~\cite{Bernal:2015xba, Heikinheimo:2016yds, Bernal:2017mqb, Heikinheimo:2017ofk, Bernal:2018hjm, Bernal:2024yhu}. The resulting DM relic abundance can differ substantially from standard freeze-in expectations~\cite{Bernal:2015ova, Bernal:2015xba, Cervantes:2024ipg}.

It is crucial to understand that the current DM abundance is influenced by both the particle physics dynamics and the cosmological history of the Universe. Given that its early evolutionary stage remains largely unknown, the conventional assumption is that the Universe was dominated by SM radiation from the conclusion of the cosmological inflation up to the point of matter-radiation equality. It is also assumed that reheating is instantaneous and occurs at a temperature much higher than the typical scales for DM production. However, this cannot be taken for granted~\cite{Allahverdi:2020bys}. In fact, cosmic reheating (that is, the era in which the inflaton transmits its energy density to SM particles, creating the SM bath)~\cite{Kofman:1994rk, Kofman:1997yn} is a continuous process that could end at a low temperature, right above the Big Bang nucleosynthesis scale.

This non-instantaneous reheating history can have a profound impact on DM production, particularly in freeze-in scenarios where DM does not reach equilibrium with the SM plasma. When self-interactions are also present, the resulting interplay between the evolving SM bath and the internal dark sector dynamics creates a thermal history far richer than that of standard freeze-in models.

In this work, we address a gap in the literature by investigating for the first time the freeze-in production of DM in the presence of both non-instantaneous reheating and dark-sector self-interactions. We numerically solve the coupled Boltzmann equations governing the DM number density, the dark sector temperature, and the evolution of the SM bath, allowing us to track the effects of both portal interactions and 2-to-3 cannibal processes in a dynamically evolving background.

Our results demonstrate that these effects can significantly shift the viable parameter space and open up new regions of phenomenological interest. In particular, we show that regions of parameter space excluded under the assumption of instantaneous reheating or negligible self-interactions can become viable when these effects are properly included.

The paper is organized as follows. In Section~\ref{sec:reh}, we discuss the dynamics of reheating while in Section~\ref{sec:DM_evol} we discuss the evolution of the dark sector. Section~\ref{sec:Z3} introduces the $\mathbb{Z}_3$ scalar DM model and the collision operators pertaining to DM production and self-interactions. Section~\ref{sec:Results} presents the core results of our numerical analysis. Finally, we summarize our conclusions in Section~\ref{sec:concl}.

\section{Evolution of the background} \label{sec:reh}
During the cosmic reheating era the energy stored in the inflaton field $\phi$ dictates the expansion of the Universe, while gradually being transferred to radiation, which we assume is composed only of SM states. The evolution of the energy densities of the inflaton $\rp$ and radiation $\rR$ determines the background for the freeze-in dynamics of the DM, where the energy density of the latter is negligible in comparison. This allows for separate determination of the background evolution first, which we then use in the following subsections for studying the evolution of the number density and temperature of the DM.

Assuming a perturbative decay of non-relativistic inflatons with a 100\% branching ratio into SM particles, and given the strength of gauge interactions their subsequent instantaneous thermalization, the evolution of the constituent energy densities can be tracked by the set of Boltzmann equations~\cite{Chung:1998rq, Giudice:2000ex}\footnote{In general, the non-perturbative and non-linear dynamics of the background during reheating is not expected to be captured by this Boltzmann approach. However, here we are interested on the last part of reheating, where the linear regime is typically a good approximation~\cite{Bassett:2005xm, Allahverdi:2010xz, Amin:2014eta, Lozanov:2019jxc, Barman:2025lvk}.}
\begin{align}
    & \frac{d\rp}{dt} + 3\, H\, \rp = -\Gamma\, \rp\,, \label{eq:BEinf}\\
    & \frac{ds}{dt} + 3\, H\, s = +\frac{\Gamma}{T}\, \rp\,, \label{eq:BEsm}
\end{align}
where the SM entropy density $s$ and its energy density are related to the SM bath temperature $T$ through
\begin{equation}
    s(T) = \frac{2\pi^2}{45}\, \gss(T)\, T^3,
\end{equation}
and
\begin{equation}
    \rR(T) = \frac{\pi^2}{30}\, \gs(T)\, T^4,
\end{equation}
with $\gss(T)$ and $\gs(T)$ being the effective number of relativistic degrees of freedom contributing to the entropy and $\rR$, respectively. In the above Boltzmann equations $\Gamma$ is the total inflaton decay width, while $H$ is the Hubble expansion rate given by
\begin{equation} \label{eq:Hubble}
    H^2 = \frac{\rp + \rR}{3\, M_P^2}\,,
\end{equation}
where $M_P \simeq 2.4 \times 10^{18}$~GeV is the reduced Planck mass.

The unknown inflaton decay width $\Gamma$ introduces an additional parameter in the theory that can be traded for the more conventional reheating temperature $\Trh$ defined as the SM temperature at the time when the energy densities of inflaton and radiation are equal; that is, $\rp(\Trh) = \rR(\Trh)$. For future reference, we also define the cosmic scale factor $\arh$ as $\arh \equiv a(\Trh)$. 

\begin{figure}[t!]
    \def\sepf{0.49}
    \centering
    \includegraphics[width=\sepf\columnwidth]{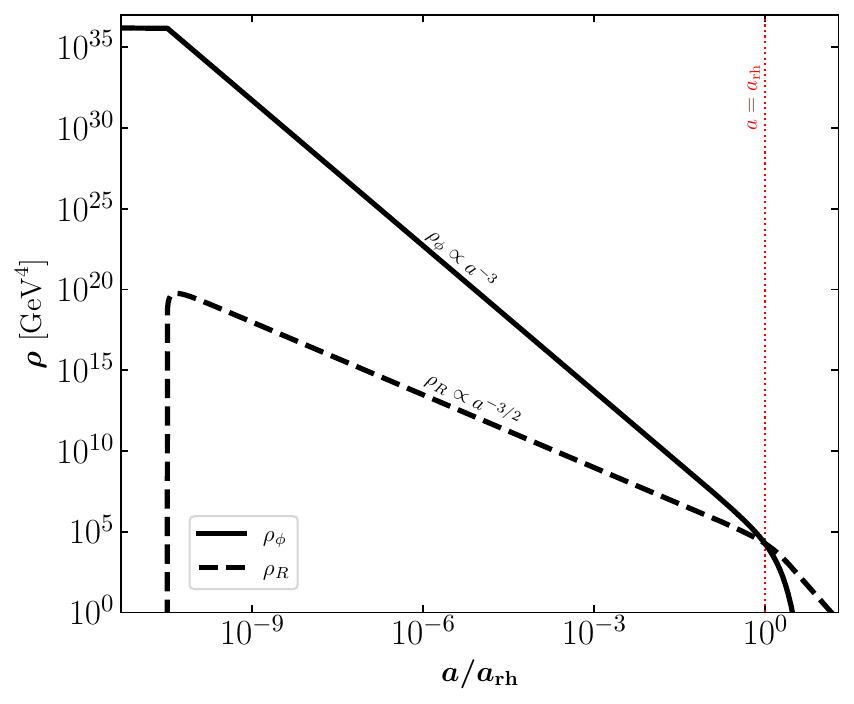}
    \includegraphics[width=\sepf\columnwidth]{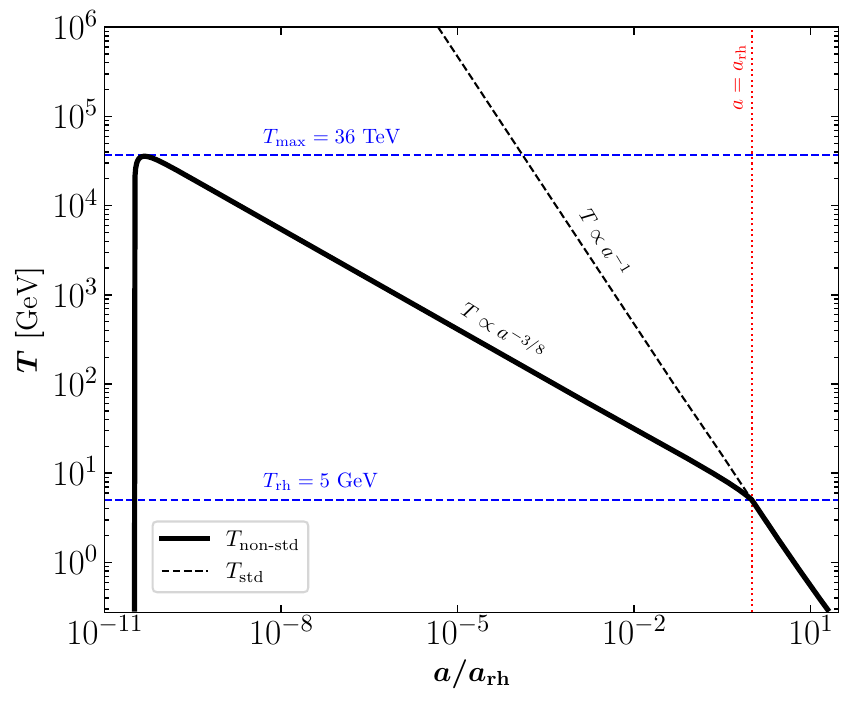}
    \caption{Evolution of the energy density of the inflaton field $\rp$ and the SM radiation $\rR$ (left) and the SM temperature $T$ (right), as a function of the cosmic scale factor $a$, for $\Trh = 5$~GeV.}
    \label{fig:energy_dens_SMtemp}
\end{figure}

The system of Boltzmann Eqs.~\eqref{eq:BEinf} and~\eqref{eq:BEsm} can be solved analytically, albeit only approximately. In fact, during reheating (that is, when $a \ll \arh$) the sink term in Eq.~\eqref{eq:BEinf} can be neglected; this allows us to integrate it and then solve Eq.~\eqref{eq:BEsm}. Alternatively, after the end of reheating ($\arh \ll a$), the source term in Eq.~\eqref{eq:BEsm} can be safely ignored. Therefore, the SM temperatures evolves as
\begin{equation}
    T(a) \simeq \Trh \times
    \begin{dcases}
        \left(\frac{\arh}{a}\right)^\frac38 & \text{ for } a \leq \arh\,,\\
        \left(\frac{\gss(\Trh)}{\gss(T)}\right)^\frac13 \frac{\arh}{a} & \text{ for } a \geq \arh\,,
    \end{dcases}
\end{equation}
while the Hubble expansion rate scales as 
\begin{equation} \label{eq:H}
    H(T) \simeq \frac{\pi}{3\, M_P}\times
    \begin{dcases}
        \sqrt{\frac{\gs(\Trh)}{10}}\, \frac{T^4}{\Trh^2} & \text{ for } T \geq \Trh\,,\\
     \sqrt{\frac{\gs(T)}{10}}\, T^2 &\text{ for } T \leq \Trh\,.
    \end{dcases}
\end{equation}
At late times, when reheating has concluded and the Universe is dominated by SM radiation, the conservation of the SM entropy density implies that $T(a) \propto 1/a$, and $H(T) \propto T^2/M_P$. An example of the evolution of the inflaton and SM radiation energy densities (left) and the SM temperature (right) as a function of the cosmic scale factor is shown in Fig.~\ref{fig:energy_dens_SMtemp}, for $\Trh = 5$~GeV. To avoid spoiling the success of Big Bang nucleosynthesis (BBN), the end of the reheating era must end at temperatures $\Trh > T_\text{BBN} \simeq 4$~MeV~\cite{Sarkar:1995dd, Kawasaki:2000en, Hannestad:2004px, Barbieri:2025moq}.

\section{Evolution of the dark sector} \label{sec:DM_evol}
On top of inflaton and the SM radiation background, the freeze-in mechanism leads to a gradual population of the dark sector, which in what follows is comprised of only one field describing the DM. Its evolution can be traced through the time dependence of the DM phase-space distribution function $f(p,t)$ that encodes the distribution of particles with momentum $p$ at a given time $t$. With the Hubble rate defined in Eq.~\eqref{eq:Hubble}, the evolution of $f$ is governed by the unintegrated Boltzmann equation
\begin{equation} \label{eq:fBE}
    \left(\partial_t - H\, \vec{p}\cdot \vec\nabla_p\right) f(p,t) = C[f]\,,
\end{equation}
where $C[f]$ denotes the collision operator encoding all interactions between DM and the surrounding thermal plasma, including self-scattering as well as elastic and inelastic scatterings with other species. The general structure of the collision term for a process $(i+j+\ldots \leftrightarrow k+\ldots)$ can be expressed as
\begin{multline}
	C [f_i] = - \int \biggl( \prod_{\alpha = j,k,\ldots} \!\!d\Pi_\alpha \biggr) \, (2\pi)^4 \, \delta^{(4)} \left( p_i\!+p_j\!\ldots\!-p_k\!-\ldots \right) \bigl\lvert \overline{\mathcal{M}} (ij \ldots \to k\ldots) \bigr\rvert^2 \\
	\times \Bigl( f_i(p_i) \, f_j(p_j) \ldots \bigl(1 \pm f_k (p_k) \bigr) \ldots - f_k (p_k) \ldots \bigl(1 \pm f_i (p_i) \bigr) \bigl(1 \pm f_j (p_j)\bigr) \ldots \Bigr),
    \label{eq:intCO}
\end{multline}
where $\lvert \overline{\mathcal{M}}  \bigr\rvert$ is the amplitude summed over the initial and final degrees of freedom and $d\Pi_\alpha$ is the Lorentz-invariant phase space measure including internal degrees of freedom and all appropriate symmetry factors. In what follows, we assume that the equilibrium distributions for both the SM and the DM states are well approximated by the Maxwell-Boltzmann form and that the factors $(1\pm f) \simeq 1$.\footnote{This approximation is motivated by the fact that in this work we focus on DM self-interactions and that the DM during freeze-in constitutes a very dilute system. Although some minor corrections are expected when considering relativistic distributions, see e.g. Ref.~\cite{Lebedev:2019ton}, these also significantly increase the complexity of the numerical integrations needed for studying the DM temperature evolutions, and as such we do not include them in this work.}

In a generic freeze-in scenario the form of the distribution $f$ is a priori unknown and needs to be solved for. This comes with a numerical challenge as the Eq.~\eqref{eq:fBE} is a non-linear partial integro-differential equation. However, in most cases one is interested only in the number density of DM, rather than the whole momentum distribution, which means it is sufficient to solve only for its zeroth moment 
\begin{equation} \label{eq:n}
    n(t) = g \int \frac{d^3p}{(2\pi)^3}\,f(p,t)\,,
\end{equation}
where $n$ is the number density and $g$ denotes the DM internal degrees of freedom. Thus integrating Eq.~\eqref{eq:fBE} over DM momentum one arrives at
\begin{equation} \label{eq:BEN}
    \frac{dN}{da}=\frac{a^2}{H}\,  g\int \frac{d^3p}{(2\pi)^3}\, C[f] \equiv \frac{a^2}{H}\, C_0\,,
\end{equation}
where in the last equality we defined the zeroth moment of the collision term and $N \equiv a^3\, n$ is the comoving number density; convenient quantity to consider during reheating when the SM entropy is not conserved. The above equation is in a closed form only if the $C_0$ can be expressed as a function of only $n$ and not $f$. In a generic freeze-in case this follows from neglecting the back reaction terms, i.e. ones describing DM annihilation and other inelastic processes involving DM in the initial state. However, in the cannibal scenario this is not possible, due to non-negligible $2\leftrightarrow 3$ processes. Nevertheless, one still can obtain a closed form set of ordinary differential equations, by extending the moments hierarchy to the next, second moment and adopting an Ansatz for the form of the distribution as a function of energy $E$ and temperature $T'$ of the DM plasma:
\begin{equation}
    f(E, T') = \frac{n}{n_\text{eq}}\, \exp\left[-\frac{E}{T'}\right],
\end{equation}
where the equilibrium number density is given by
\begin{equation}
    n_\text{eq}(T') = \frac{g}{2 \pi^2} \int dE\, p\, E\, \exp\left[-\frac{E}{T'}\right] = \frac{g\, m^2\, T'}{2 \pi^2}\, K_2\left(\frac{m}{T'}\right).
\end{equation}
This assumption is motivated by the fact that generically existence of efficient cannibal processes requires also efficient elastic self-scatterings between the DM particles, which in turn drive the distribution function to an equilibrium shape. Note however, that the DM temperature does not have to be the same as the SM plasma one, and in fact considering both moments of Eq.~\eqref{eq:fBE} one arrives at two equations: for DM number density $n$ and its temperature $T'$. The latter is obtained by integrating both sides of Eq.~\eqref{eq:fBE} over $g\, (2\pi)^{-3} \int d^3p\, p^2/E$
\begin{equation} \label{eq:Tp0}
    \frac{d(3\, n\, T')}{dt} - H\, g \int\frac{d^3p}{(2\pi)^3}\, \frac{p^2}{E}\, \vec{p} \cdot \vec\nabla_p f  = g \int \frac{d^3p}{(2\pi)^3}\, \frac{p^2}{E}\, C[f]\, \equiv C_2\,, 
\end{equation}
where on the left hand side we introduced the DM effective `temperature' through~\cite{Bringmann:2006mu, Binder:2021bmg}
\begin{equation}
    T' = \frac{g}{3\, n} \int \frac{d^3p}{(2\pi)^3}\, \frac{p^2}{E}\, f(p,t)\,,
\end{equation}
relating it to the second moment of the distribution. $T'$ coincides with the actual physical temperature if the DM is in kinetic equilibrium, i.e. when the distribution function follows the Maxwell-Boltzmann form, albeit with arbitrary chemical potential. For generic, non-thermal distributions the notion of temperature is ill-defined, however $T'$ retains its meaning as a velocity dispersion parameterizing the typical kinetic energy of the DM particles.

Finally, noticing that
\begin{equation}
    \frac{dn}{dt} = C_0 - 3\, H\, \frac{N}{a^3}
\end{equation}
and
\begin{align}
    &g \int\frac{d^3 p}{(2\pi)^3}\, \frac{p^2}{E}\, \vec{p} \cdot \vec\nabla_p f =  g \int \frac{d^3p}{(2\pi)^3} \left(\frac{p^4}{E^3} - 5\, \frac{p^2}{E}\right) f = n \left\langle\frac{p^4}{E^3}\right\rangle -15\, n\, T', 
\end{align}
Eq.~\eqref{eq:Tp0} can be conveniently rewritten as
\begin{equation} \label{eq:cBE}
    \frac{dT'}{da} = - \frac{2\,  T'}{a} + \frac{a^2}{3\, H\, N}\, C_2 - \frac{a^2\, T'}{H\, N}\, C_0 + \frac{1}{3\, a} \left\langle\frac{p^4}{E^3}\right\rangle.
\end{equation}
In this formulation, given $C_0$ and $C_2$ that follow from the interactions present in a given model, it is possible to numerically solve the equations for the number density and temperature of DM (Eqs.~\eqref{eq:BEN} and~\eqref{eq:cBE}, respectively) in the background defined by Eqs.~\eqref{eq:BEinf} and~\eqref{eq:BEsm}.

\section{\boldmath $\mathbb{Z}_3$ scalar dark matter} \label{sec:Z3}
\subsection{The model}
We consider a minimal extension of the SM with a complex DM candidate denoted as $S$, neutral under the SM gauge group but charged under a hidden global $\mathbb{Z}_3$ symmetry, transforming as $S\to e^{\frac{2\pi}{3}\, i}\, S$. It naturally couples to the SM Higgs doublet $\widetilde H$ via the Higgs-portal interaction with the potential
\begin{equation} \label{eq:VH}
    V_\text{HP}(\widetilde H,S) = \lhs\, \vert \widetilde H\vert^2\, \vert S\vert^2\,,
\end{equation}
where $\lhs$ is the corresponding dimensionless coupling. Furthermore, in addition to the mass term, the scalar field $S$ can also have trilinear and quartic self-interactions of the form
\begin{equation}
    V_s(S) =\mu_s^2\, \vert S\vert^2 + \frac{g_s}{3!} \left(S^3+(S^*)^3\right) + \frac{\ls}{4}\, \vert S\vert^4\,.
\end{equation}
The stability of the potential is ensured when $k \equiv g_s^2 / (3\, \ls\, \mu_s^2) < 8/3$~\cite{Cervantes:2024ipg}. The physical DM mass $m^2 = \mu_s^2 + \lhs\, v^2/2$, where $v = 246$~GeV is the vacuum expectation value (VEV) of the Higgs field.

A $\mathbb{Z}_3$-stable scalar was initially introduced in the context of neutrino physics~\cite{Ma:2007gq}. As a potential WIMP candidate, its phenomenology was first explored in Ref.~\cite{Belanger:2012zr} and later expanded in Ref.~\cite{Hektor:2019ote} in the context of DM annihilations, but also for DM semi-annihilations~\cite{DEramo:2010keq}. However, the $\mathbb{Z}_3$ symmetry also allows for cubic (and quartic) self-coupling, leading to cannibalistic interactions of the form $3 \leftrightarrow 2$, where three DM particles are annihilated into two~\cite{Carlson:1992fn, Hochberg:2014dra, Ko:2014nha, Choi:2015bya, Bernal:2015bla, Bernal:2015lbl}. The larger the cubic self-coupling, the stronger the $3 \leftrightarrow 2$ processes compared to the $4 \leftrightarrow 2$ one~\cite{Bernal:2015xba, Heikinheimo:2016yds, Bernal:2017mqb, Heikinheimo:2017ofk, Bernal:2018hjm}, since the latter involves an additional power of the quartic coupling. In this analysis, we set $k = 2$; this is motivated by the choice to focus on $3\leftrightarrow 2$ over $4\leftrightarrow 2$ reactions and the need to reduce the parameter space of the model while retaining the parameterization of the strength of self-interactions through the coupling $\ls$.

\subsection{Freeze-in production} \label{sec:FI_reh}
The portal interaction in Eq.~\eqref{eq:VH} gives rise to three types of production processes: direct Higgs decay ($h\to S^* S$), direct $2\leftrightarrow 2$ ($hh\to S^*S$) and Higgs-mediated production from other SM states populating the plasma. Out of these, whenever kinematically possible and as long as Higgs bosons are abundant in the plasma, the decay mode dominates. However, if the reheating temperature is low, contributions from fermionic and hadronic $2\leftrightarrow 2$ processes cannot be neglected~\cite{Lebedev:2024mbj}. Nevertheless, in both the high and the low $\Trh$ cases, the production that occurs before the electroweak phase transition is much smaller than the two other modes, and therefore we set the initial condition at $a_I = a_\text{ewpt}$ in this work, where $T(a_\text{ewpt}) = 150$~GeV~\cite{Heeba:2018wtf}. Note that we also fix the inflationary scale such that the maximal temperature reached by the SM plasma during reheating is much higher than $T_\text{ewpt}$, ensuring that all SM species remain in thermal equilibrium at all times, that is, during and after reheating.

Freeze-in production gradually populates the dark sector, but by itself does not make it thermalize. This is achieved through self-interactions, both elastic and inelastic. In particular, due to $\mathbb{Z}_3$ charge conservation, the only $2 \leftrightarrow 3$ number-changing reactions allowed within the dark sector are $S^*S \to SSS$ and $SS \to S^*S^*S$, together with their complex conjugates. These processes trade excess kinetic energy to a number of $S$ states, boosting the freeze-in production~\cite{Bernal:2020gzm} and later in the evolution, when the temperature drops below $m$ conversely lead to cannibalization and heating of the dark sector with respect to the SM plasma~\cite{Carlson:1992fn}.

Finally, note that we assume that the dark sector has no initial asymmetry and that there is no CP violation in the model. This leads to a total number density of DM particles $n$ being simply the sum of the number densities of $S$ and $S^*$; that is, $n = n_s + n_{s^*} = 2\, n_s$.

\subsubsection{Production from the Higgs decay} \label{subsec:FI_reh1}
The dilution of the DM energy density due to the rapid expansion during reheating ceases once the Universe reaches the reheating temperature $\Trh$. When $\Trh$ is of the order of a few GeV or higher, the dominant DM production mechanism is the Higgs decay process $h \to S^*S$, which contributes to the comoving number density equation as given in Ref.~\cite{Cervantes:2024ipg}, 
\begin{equation}
    C_0^{h\to S^*S}=\frac{\lhs^2\, v^2\, m_h}{16\pi^3}\sqrt{1-\frac{4m^2}{m_h^2}}\, T\, K_{1}(m_h/T)\,,
\end{equation} 
with $m_h$ being the Higgs mass. Meanwhile, the contribution to the temperature evolution equation is approximately (under the assumption $m \ll m_h$)
\begin{equation}\label{eq:FI_2}
    C^{h\to S^*S}_2\simeq \frac{\lhs^2\, v^2\, m_h^2}{32\pi^3}\sqrt{1-\frac{4m^2}{m_h^2}}\, T\, K_2(m_h/T)\,,
\end{equation}  
where in the above equations $K_n(x)$ represents the modified Bessel function of the second kind of order $n$.

\subsubsection{Production from fermionic and hadronic collisions} \label{subsec:FI_reh2}
Alternatively, if reheating ends at temperatures in the MeV range, Higgs-induced production occurs during a period of significant dilution. To compensate for the diluted DM, an enhanced production is required, which can originate from the primordial plasma. This requires higher values of $\lhs$, increasing the production rates of both fermionic and hadronic SM states. The latter becomes available only after the QCD phase transition. Note that higher values of $\lhs$ can lead to kinetic and chemical equilibrium between the dark and SM sectors, resulting in a transition from DM being a FIMP to a WIMP~\cite{Silva-Malpartida:2023yks, Belanger:2024yoj}. If the reheating temperature is very low, the interaction rates will decouple before the reheating ends. In such cases, proper treatment of temperature evolution is required to account for kinetic decoupling accurately. 

The DM temperature is affected by both annihilations and elastic scatterings with the thermal bath particles. The annihilation of the SM states that produce DM contributes to both $C_0$ and $C_2$. For example, $C_0^{S^*S\leftrightarrow \bar f f}$ can be expressed in terms of the thermal average\footnote{Here $\text{v}$ is the relative velocity of $S^*$ and $S$ prior the collision and not the SM VEV as previously defined.} $\braket{\sigma_{S^*S\to \bar f f} \text{v}}$ detailed in Ref.~\cite{Edsjo:1997bg}. Furthermore, $C_2^{S^*S\leftrightarrow \bar f f}$ is expressed in terms of the temperature-weighted average $\braket{\sigma_{S^*S\to \bar f f} \text{v}}_2$, which is defined and can be found in Ref.~\cite{Binder:2017rgn}. These contributions describe interactions with leptons and quarks in the deconfined phase, which we assume to be the case prior to the QCD phase transition, which we assume to be instantaneous at $T_\text{QCD}=200$~MeV. After the QCD transition ($T<T_\text{QCD}$) the quarks are confined into hadrons, hence we use the decay width $\Gamma_\text{hadrons}$ presented in Ref.~\cite{Winkler:2018qyg}. Their contribution to $C_0$ is
\begin{equation}\label{eq:C0hadrons}
    C_0^{S^*S\to \text{hadrons}} =-\int\frac{d^3p}{(2\pi)^3} \frac{d^3\tilde p}{(2\pi)^3}\sigma \text{v}_{S^*S\to \text{hadrons}}f_s(p)f_s(\tilde p)\,,
\end{equation}
where
\begin{equation}
    \sigma\,\text{v}_{S^*S\to \text{hadrons}} \equiv \frac{\lhs^2v^2}{\sqrt{s}}\frac{1}{(s-m_h^2)^2 + m_h^2\Gamma_h(m_h)^2}\Gamma_{\text{hadrons}}(\sqrt{s})\,,
\end{equation}
with $\Gamma_h(m_h) = 4.042$~MeV~\cite{Djouadi:2005gi}.
The respective second-moment integral is
\begin{equation}
    C_2^{S^*S\to \text{hadrons}} = -\int\frac{d^3p}{(2\pi)^3}\, \frac{d^3\tilde p}{(2\pi)^3}\, \frac{p^2}{E}\, \sigma\text{v}_{S^* S \to \text{hadrons}}\, f_s(p)\, f_s(\tilde p)\,.
\end{equation}
Lastly, elastic collisions of DM with SM states impact the temperature evolution driving $T'$ to $T$. Right after the EWPT, elastic collisions with the Higgs boson can be estimated as~\cite{Cervantes:2024ipg}
\begin{equation}
    C_2^{S h\leftrightarrow Sh} \simeq \lhs^2\, \frac{N}{N_\text{eq}}\, \frac{1}{128\pi^5}\, T'^2\, T^2 \left(T - T'\right) e^{-\frac{m}{T'}}\, e^{-\frac{m_h}{T}},
\end{equation}
and they rapidly decouple due to the exponential factor $\text{exp}{(-m_h/T)}$. The second source of kinetic equilibration is the scattering with fermions, which we estimate as
\begin{equation} \label{eq:C2_el_scatter_ferm}
    C_2^{S f\leftrightarrow Sf} \simeq \frac{N}{N_\text{eq}}\, \frac{\lhs^2\, m_f^2\, N_c}{64 \pi^5\, m_h^4}\, T'^2\, T^2 \left(T - T'\right) \left(m_f^2 + 3\, T\, T'\right) e^{-\frac{m}{T'}}\, e^{-\frac{m_f}{T}},
\end{equation}
where $N_c=1$ for leptons and $N_c=3$ for quarks; cf Appendix~\ref{ap:A} for a detailed derivation.

\subsubsection{Self-interactions collision integrals} \label{subsec:FI_reh3}
Since we treat $S^*$ and $S$ as different states that contribute equally to the system of Boltzmann equations, it is important to keep track of the collision operator of each state. For the reaction $S^*S\leftrightarrow SSS$ and the particle $S^*$ the collision operator is
\begin{equation}
    C_{S^*S\leftrightarrow SSS}[S^*]=\frac{1}{2E_{s^*}}\frac{1}{3!}\int \vert\tilde{\mathcal{M}}\vert^2 d\Pi_2d\Pi_3d\Pi_4d\Pi_5\left(f_3f_4f_5 
        -f_{s}  f_2  \right),
\end{equation}
whereas the collision operator for $S$ is
\begin{equation}
    C_{S^*S\leftrightarrow SSS}[S] = \frac{1}{2E_{s}}\int\vert\tilde{\mathcal{M}}\vert^2 d\Pi_1d\Pi_3d\Pi_4d\Pi_5\left(-\frac{1}{3!}f_s f_1 + \frac{1}{3!}f_3f_4f_5 +\frac{1}{2!}f_1f_2-\frac{1}{2!}f_s f_4 f_5\right).
\end{equation}
where $\vert\tilde{\mathcal{M}}\vert^2=(2\pi)^4\delta^{(4)}\left(\sum_f p_f - \sum_i p_i\right)\vert\mathcal{M}_{3\leftrightarrow 2}\vert^2$. The reaction $S^*S^*S\leftrightarrow SS $, as well as the computation of $C_0$ and $C_2$ are discussed in Ref.~\cite{Cervantes:2024ipg}. 

\section{Results} \label{sec:Results}
The interplay between freeze-in production through portal coupling $\lhs$, number-changing self-interactions controlled by $\ls$, and the evolving Hubble rate during and after reheating creates a complex dynamics. To facilitate the discussion of our main results, we begin this section by exploring two benchmark cases that illustrate the interrelation of various processes. They correspond to representative cases that fit the observed DM relic abundance, where the freeze-out of the DM self-interactions occurs after or during reheating. We then try to generalize our results by performing a large scan over the relevant parameters. 

In order for the different interaction rates to be effective, they must compete with the Hubble rate, which scales as $T^4/(M_P \Trh^2$) during reheating and $T^2/M_P$ afterward; see Eq.~\eqref{eq:H}. At high temperatures, DM production is primarily driven by Higgs decays, but the rate rapidly declines due to the Boltzmann suppression as temperature decreases. However, most of the DM produced is diluted because reheating takes place; its impact is more severe for smaller values of $\Trh$. Hence, DM production via the annihilation of light SM states through an off-shell Higgs becomes relevant for small $\Trh$ values, albeit with a lower rate compared to the Higgs decay, necessitating a higher portal value $\lhs$. The inverse process of DM annihilating to SM particles remains suppressed because of the insufficient number density of DM, as expected in this FIMP scenario. Self-interactions drop as $m^{-5}$, making them impactful only in the low DM-mass range. Initially, when the density of the DM number is small, but enough kinetic energy is available, the $2 \to 3$ process is the only relevant. When DM becomes abundant enough, the importance of $3 \to 2$ process grows, eventually enforcing equilibrium and resulting in a characteristic temperature $T'$ of the DM bath.

\subsection{Dark freeze-out after reheating} \label{subsec:trh1GeV}
\begin{figure}
    \def\sepf{0.496}
    \centering
    \includegraphics[width=\sepf\columnwidth]{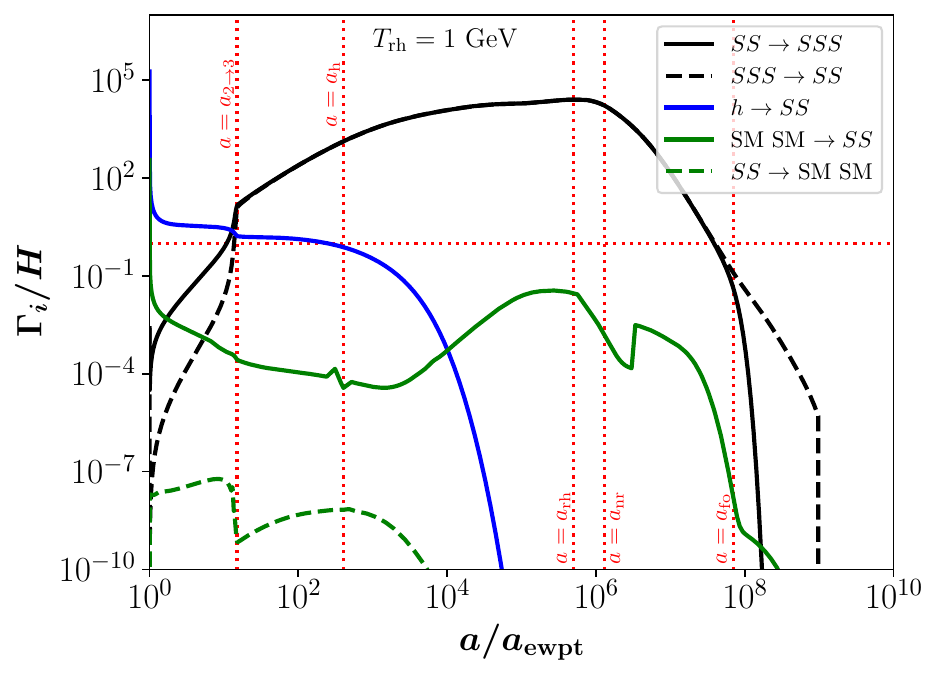}
    \includegraphics[width=\sepf\columnwidth]{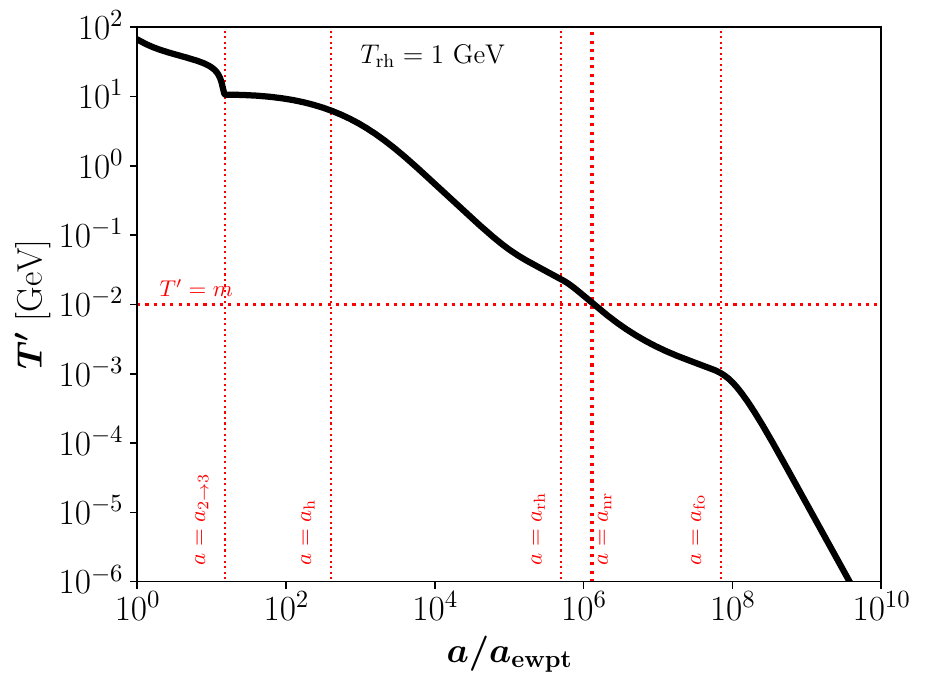}
    \includegraphics[width=\sepf\columnwidth]{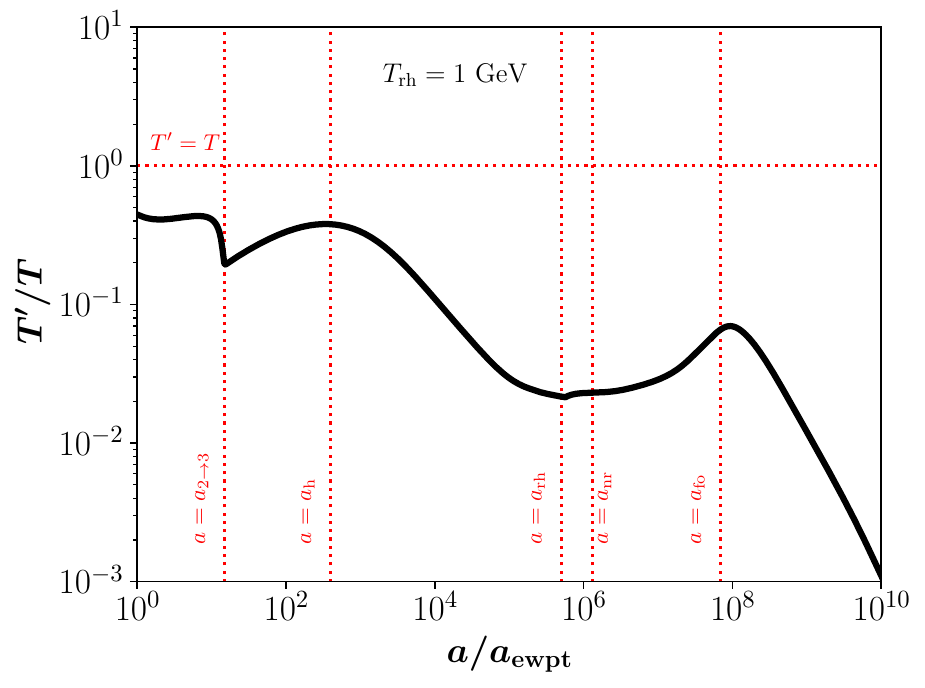}
    \includegraphics[width=\sepf\columnwidth]{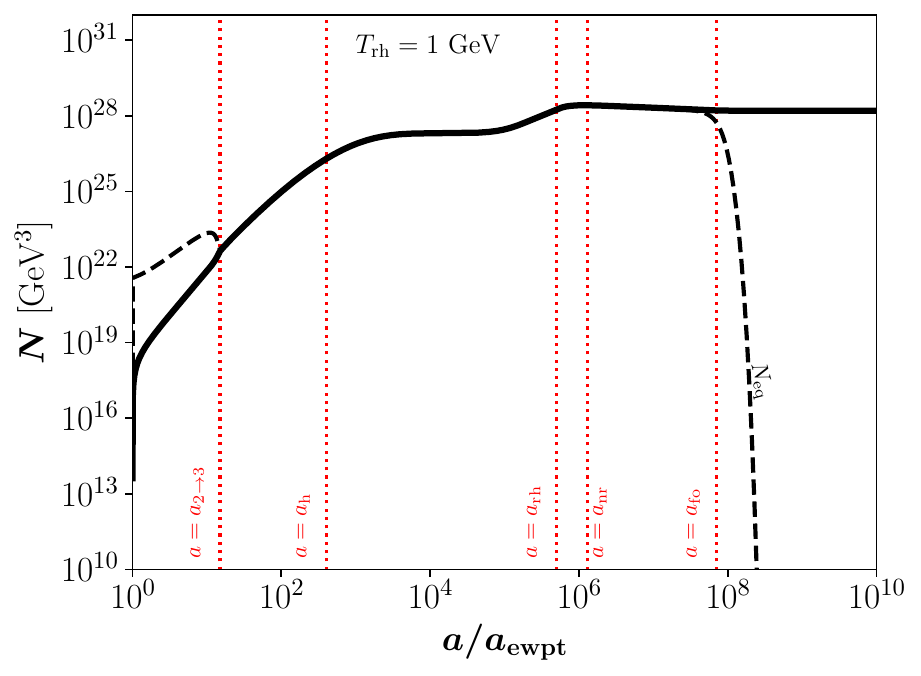}
    \caption{Results for $\Trh = 1$~GeV, setting $m = 10$~MeV, $\lhs = 1.8 \times 10^{-7}$ and $\ls =0.01$. Top left: Ratio of the different interaction over the Hubble rate. Top right: Temperature evolution of the dark sector. Bottom left: Ratio of the dark sector and SM temperature. Bottom right: Evolution of the co-moving number density.}
    \label{fig:Trh1GeV}
\end{figure}

Figure~\ref{fig:Trh1GeV} presents the results for $\lhs = 1.8 \times 10^{-7}$ and $\Trh = 1$~GeV, setting the DM mass to $m = 10$~MeV and self-coupling to $\ls = 0.01$. Here, DM is mainly sourced from Higgs decays. The top left panel illustrates the ratio of various interaction rates to the Hubble parameter, while the top right panel depicts the evolution of the dark sector temperature. The bottom left panel shows the temperature ratio between the dark sector and the SM, and the bottom right panel displays the evolution of the comoving number density of the DM.

Initially, Higgs decays (solid blue curve in the top-left panel) dominate and establish the initial DM abundance. Consequently, the DM temperature is $T'\sim m_h/2$ when the SM temperature is comparable to the Higgs mass $m_h$. As the DM number density increases, the $2 \to 3$ self-production process (thick black curve) becomes significant, leading to some cooling of the DM sector due to the exchange of kinetic energy for the number of particles. Eventually, this process surpasses both the Higgs decay and the Hubble rates, as indicated by the first vertical dotted line at $a = a_{2 \to 3}$. This marks a sudden drop in the DM temperature. However, as DM becomes sufficiently abundant, the inverse $3 \to 2$ self-annihilation process (dashed black curve) becomes competitive, bringing number-changing processes into equilibrium: DM reaches chemical equilibrium with itself, that is, $N = N_\text{eq}$. Higgs-mediated DM production continues until its decay rate falls below the Hubble rate, which is indicated by the second vertical dotted red line at $a = a_\text{h}$. Throughout this period, SM annihilations (solid green line) remain inefficient.

At this stage, DM is effectively source-free and relativistic, causing its temperature to scale as $a^{-1}$. Although subdominant, SM annihilations contribute some DM production near the end of reheating, marked by the third vertical dotted red line at $a = \arh$. The contours of the ratio of various rates over Hubble change slope at this line as the slope of Hubble itself changes from $a^{-3/2}$ to $a^{-2}$.  When the DM temperature reaches approximately $T' \approx m$ at $a = a_\text{nr}$, the DM transitions to the non-relativistic regime, leading to a Boltzmann suppression of its number density. Due to entropy conservation in the dark sector, the temperature follows a logarithmic scaling, approximately $T' \sim 1/\log(a^3)$~\cite{Carlson:1992fn}. This results in an increase of $T'$ with respect to $T$, characteristic of cannibalization processes. Finally, when DM undergoes a dark freeze-out of self-annihilations (last vertical dotted red line at $a = a_\text{fo}$), its temperature begins to scale as $a^{-2}$, as it is a non-relativistic particle out of equilibrium. Meanwhile, the SM temperature evolves as $T\propto a^{-3/8}$ during reheating and as $T\propto a^{-1}$ afterward. This explains the behavior of the temperature ratio between the dark sector and the SM, as shown in the bottom left panel of Fig.~\ref{fig:Trh1GeV}.

The different phases of DM production are also evident in the bottom right panel, which illustrates the evolution of the comoving number density ($N \equiv n\, a^3$), represented by the thick black line. For reference, the equilibrium comoving number density ($N_\text{eq} \equiv n_\text{eq}\, a^3$) is shown as a dashed black line.

\subsection{Dark freeze-out during reheating} \label{subsec:trh10MeV}
\begin{figure}
    \def\sepf{0.496}
    \centering
    \includegraphics[width=\sepf\columnwidth]{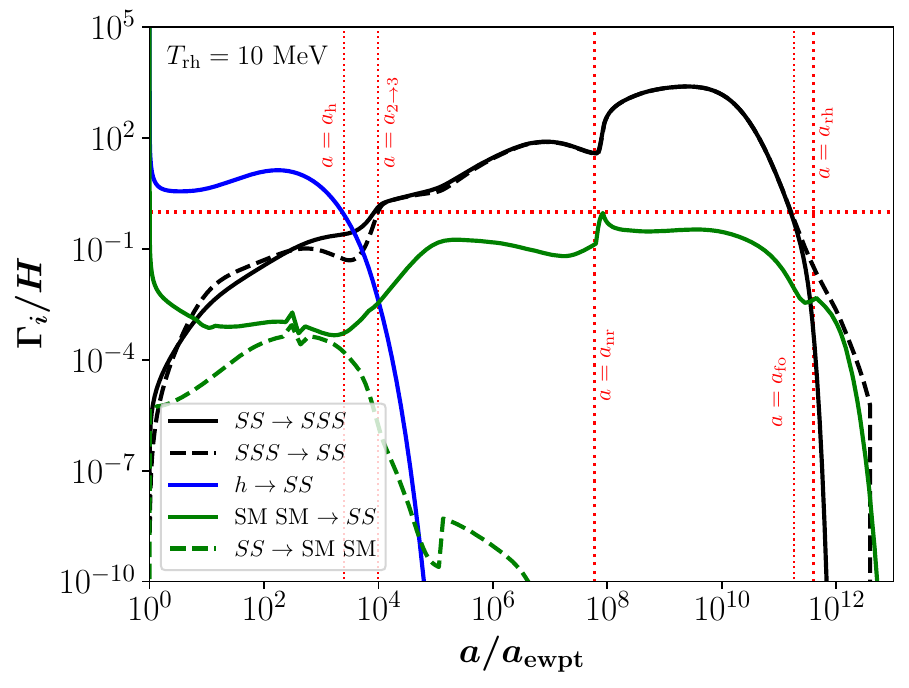}
    \includegraphics[width=\sepf\columnwidth]{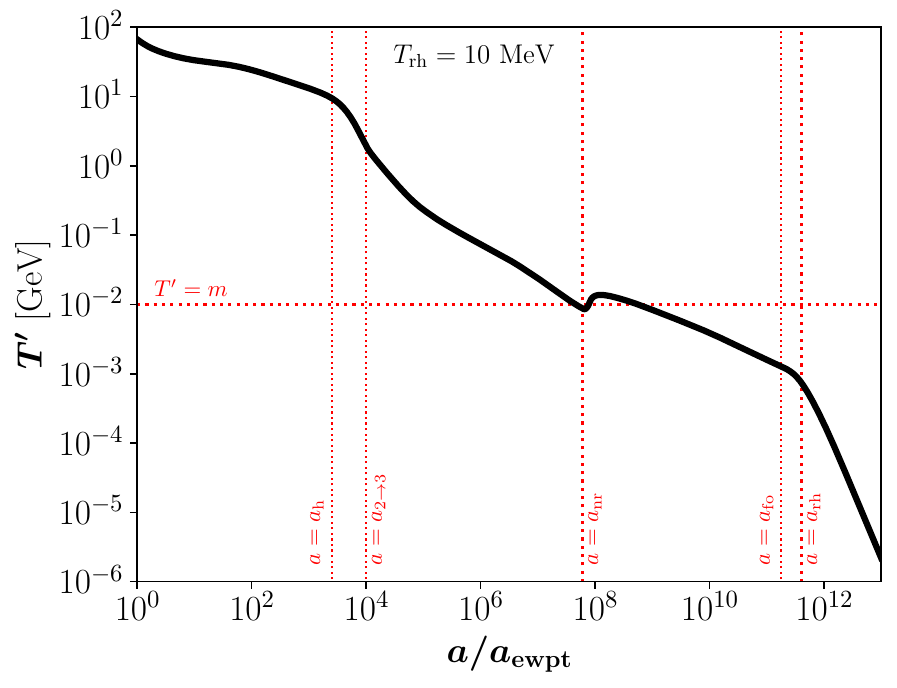}
    \includegraphics[width=\sepf\columnwidth]{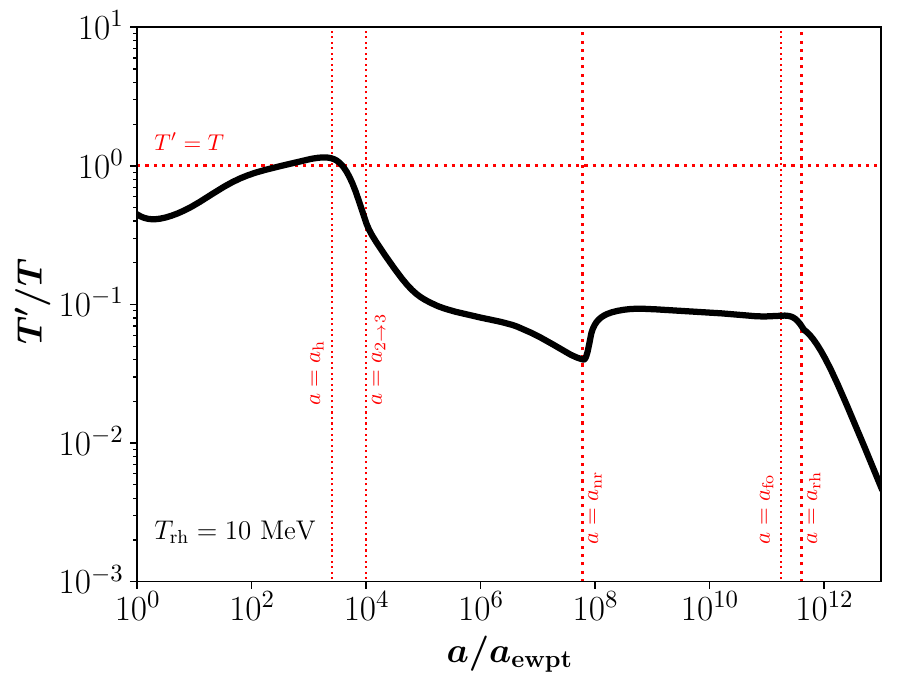}
    \includegraphics[width=\sepf\columnwidth]{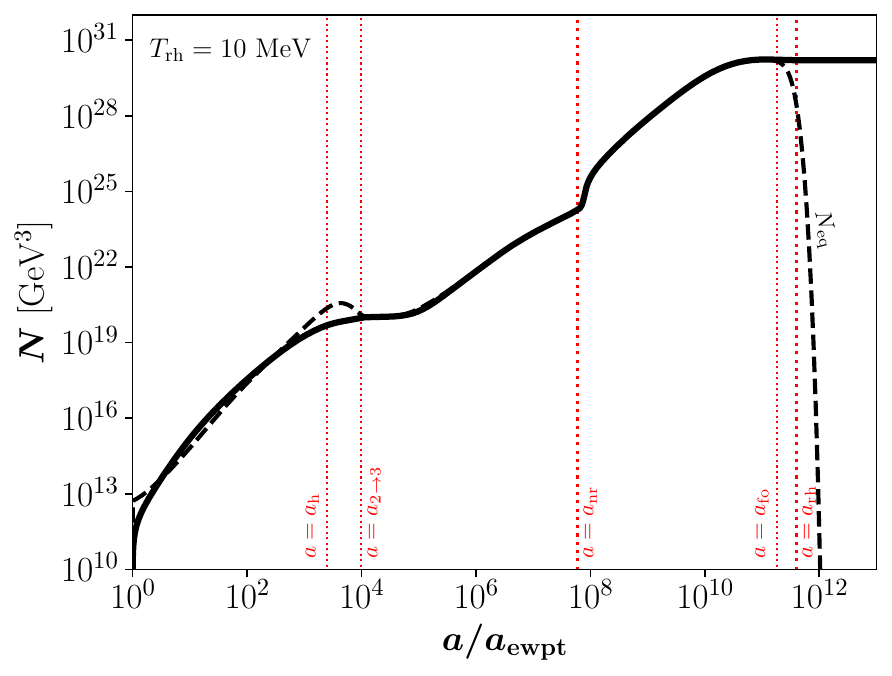}
    \caption{Same as Fig.~\ref{fig:Trh1GeV}, but showing results for $\Trh = 10$~MeV, setting $\lhs = 4.5 \times 10^{-4}$.}
    \label{fig:Trh10MeV}
\end{figure}

We present the results for a lower reheating temperature $\Trh = 10$~MeV in Fig.~\ref{fig:Trh10MeV}, sticking to the same $m$ and $\ls$ as in Section~\ref{subsec:trh1GeV}, but taking $\lhs = 4.5 \times 10^{-4}$. Due to the very low reheating temperature, the entropy dilution is significant, necessitating a much larger portal coupling $\lhs$ compared to the case of $\Trh = 1$~GeV. Consequently, self-interactions are initially less efficient than Higgs decays and the Hubble expansion, as observed in the top left panel. Higgs decays serve as the primary source of DM production, leading to a gradual increase in the temperature ratio of the dark sector and the SM, as illustrated in the bottom-left panel. This trend persists until the Higgs decay rate drops below the Hubble rate, marked by the first vertical red dotted line at $a = a_\text{h}$.

Following this point, the $2 \to 3$ self-production process briefly becomes efficient, causing a sudden drop in the DM temperature and consequently a decrease in the temperature ratio. It ends soon after, shown by the second red-dotted vertical line at $a_{2\to 3}$, when $3 \to 2$ self-annihilations become efficient and bring the number-changing processes into equilibrium. Meanwhile, SM annihilations ($2\,\text{SM} \to 2\,S$) gradually gain significance, preventing the DM from ever being fully source-free. This results in a slow decline in the DM temperature and a nearly constant temperature ratio between the DM and the SM.

As evolution continues, the DM transitions to the non-relativistic regime at $a = a_\text{nr}$ (third red dotted line), after which its number density hits the Boltzmann suppression, and the DM temperature starts to follow the typical logarithmic dependence on $a$. However, around the same time, SM annihilation processes receive a slight enhancement at the hadronization scale (characterized by the sharp increase in the self-interaction and SM annihilation rates), providing an additional source of DM production. This leads to a small increase in the DM temperature and the temperature ratio, making it a complicated function of $a$, distinguishing this scenario from the $\Trh = 1$~GeV case, where temperature evolution follows a simpler trend. 

Subsequently, the temperature ratio becomes constant until freeze-out occurs, represented by the fourth vertical red dotted line at $a_\text{fo}$. The reheating phase ends shortly after freeze-out, as indicated by the final vertical red line at $a = \arh$. The different stages of DM production are also evident in the bottom right panel, which illustrates the evolution of the comoving number density $N$. Furthermore, it is also clear that a significantly larger population of DM is required compared to the $\Trh = 1$~GeV scenario to compensate for the strong entropy dilution.\footnote{One important point to consider here is that because we set $\ls$ to $10^{-2}$ for the rates in Sections~\ref{subsec:trh1GeV} and~\ref{subsec:trh10MeV}, $2 \rightarrow 3$ interactions never impact the DM production significantly. The situation changes when $\ls$ approaches unity, as we shall see in Section~\ref{subsec:param-regions} and Fig.~\ref{fig:ls_vs_relic}. It will change the behavior of the DM temperature and comoving number density $N$ until $3 \rightarrow 2$ self-interactions catch up.}
    
\subsection{Parameter regions} \label{subsec:param-regions}
\begin{figure}[t!]
    \centering
    \includegraphics[width=0.49\textwidth]{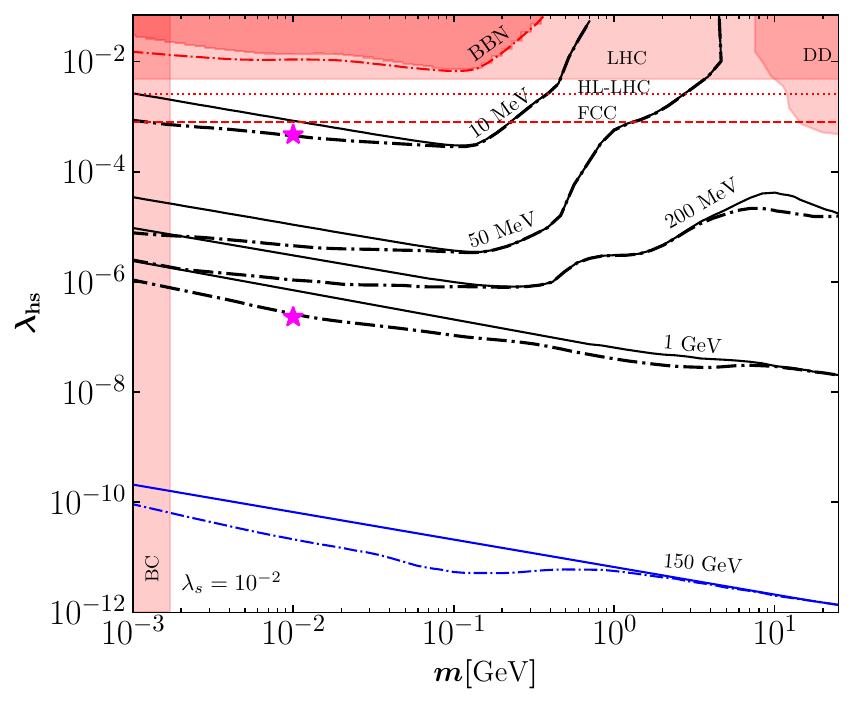}
    \includegraphics[width=0.49\textwidth]{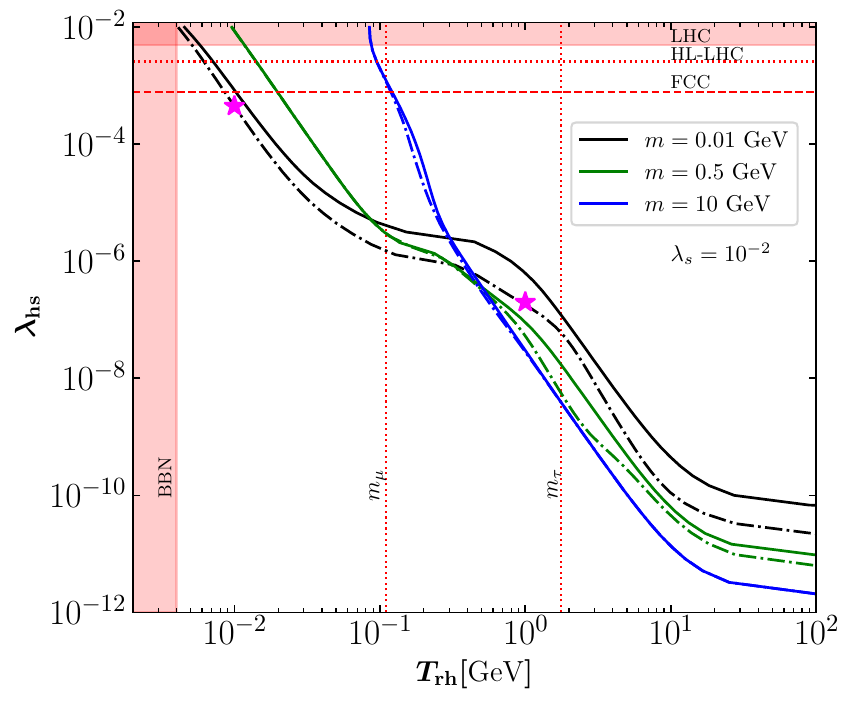}
    \caption{Left panel: Parameter space that fits the observed DM relic in the plane $[m,\, \lhs]$ for different $\Trh$. Right panel: Parameter space that fits the observed DM relic in the plane $[\Trh,\, \lhs]$ for different values of $m$. In the two panels, the dot-dashed lines correspond to $\ls = 10^{-2}$, while the solid lines correspond to the case without self interactions ($\ls = g_s = 0$).}
    \label{fig:sol_lhs_ms_Trh_FIMPs}
\end{figure}

Now that characteristic benchmarks have been discussed, we proceed to present scans of the model parameters. In the left panel of Fig.~\ref{fig:sol_lhs_ms_Trh_FIMPs}, we show contours of different $\Trh$ values in the $[m,\, \lhs]$ plane. The dot-dashed lines correspond to scenarios that include DM self-interactions for $\ls = 10^{-2}$, whereas the solid lines represent cases without self-interactions $\ls = g_s = 0$. The two magenta stars represent the two benchmark cases studied in detail in Sections~\ref{subsec:trh1GeV} and~\ref{subsec:trh10MeV}.

For very large $\Trh$ values, generally larger than the Higgs and DM masses, and specifically $\Trh = 150$~GeV in our case, the contour (shown in blue) aligns with the expectations of a DM produced after reheating, during the standard radiation-dominated era. If self-interactions of the DM are absent, the required portal coupling is in the range $\lhs \sim \mathcal{O}(10^{-11})$, characteristic of infrared freeze-in production. The enhancement in DM production via the $2 \to 3$ self-interaction process, induced by the presence of $\ls$, allows a smaller portal coupling $\lhs$, clearly visible from the thick black line. As expected, the influence of self-interactions diminishes for larger DM masses. 

As the reheating temperature decreases, the bulk of DM production occurs during the reheating era, and therefore the effect of entropy dilution necessitates a larger portal coupling $\lhs$. For intermediate $\Trh$ values (corresponding to the 200~MeV), self-interactions become significant even for DM masses in the range of $\sim [2,\, 20]$~GeV. This can be attributed to the fact that annihilations involving heavy fermions ($\tau$, $c$, and $b$) become efficient in this mass range due to their relatively large Yukawa couplings, leading to an increased abundance of DM and consequently enhancing self-interaction rates. 

For $\Trh$ values between 10~MeV and 100~MeV, entropy dilution is so pronounced that the DM produced by Higgs decays and heavy fermion annihilations is largely diluted. As a result, DM production becomes highly dependent on annihilations involving muons. This explains the sharp exponential increase in the required portal coupling once the DM mass exceeds the muon threshold $m \sim 100$~MeV, as it must compensate for the rapidly decreasing muon number density.  

Self-interactions are constrained by astrophysical observations of DM behavior in galaxies and galaxy clusters. A particularly strong constraint arises from the merging galaxy cluster 1E 0657-56 (Bullet cluster)~\cite{Randall:2008ppe}, which limits the DM self-scattering cross section per unit mass to be
\begin{equation}
    \frac{\sigma_T}{m} < 1\,\text{cm}^2/\text{g}\,
\end{equation}
at a typical DM velocity of $v \simeq 10^{-4}$. Here, the transfer cross section is defined as
\begin{equation}\label{eq:sigmaT}
    \sigma_T = \int d\Omega\, (1 - \cos\alpha)\, \frac{d\sigma_{2\to 2}}{d\Omega}\,,
\end{equation}
with $\sigma_{2\to 2}$ corresponding to the self-scattering process $SS \to SS$ and $SS^* \to SS^*$. This imposes a lower bound of approximately 1~MeV on the DM mass in the case $\ls = 10^{-2}$ and is shown by the vertical red region in the left panel (labeled BC).

The process $h \to S^* S$ gives a contribution to the decay width of the Higgs, given by
\begin{equation}
    \Gamma_{h\to S^*S} = \frac{\lhs^2\, v^2}{32\pi\, m_h} \sqrt{1-\frac{4m^2}{m_h^2}}\,,
\end{equation}  
which then gets constrained by the branching ratio of invisible decay width of Higgs at the LHC ($\text{BR}_{\text{inv}} \leq 10 \%$~\cite{ATLAS:2023tkt, CMS:2023sdw}) and places an upper limit on the portal coupling $\lhs$, depicted as the red-shaded horizontal region. Future projections from the HL-LHC ($\text{BR}_{\text{inv}} \leq 1.9\%$) and FCC ($\text{BR}_{\text{inv}} \leq 0.2\%$) on the invisible Higgs decay are also indicated by the dotted and dashed horizontal red lines, respectively~\cite{Dawson:2022zbb}.

Since reheating must be completed before the onset of BBN, $\Trh \lesssim 4$~MeV~\cite{Sarkar:1995dd, Kawasaki:2000en, Hannestad:2004px, Barbieri:2025moq} are ignored. This exclusion is represented by the red region at the top-left, which is already ruled out by LHC constraints. Lastly, direct detection experiments probing DM-nucleon scattering (LZ~\cite{LZ:2024zvo} and Xenon1T~\cite{XENON:2018voc}) impose upper limits on the portal coupling for DM masses around 1~GeV or higher, depicted by the upper-right red region (labeled as DD).
In contrast, constraints from DM-electron scattering are very weak because of the small Yukawa coupling of the electron to Higgs, leading to no significant bounds.

The right panel of Fig.~\ref{fig:sol_lhs_ms_Trh_FIMPs} shows contours of different DM masses in the $[\Trh,\, \lhs]$ plane. The underlying explanation follows from the previous plot, highlighting the dependence on muons and taus, along with charm and bottom quarks, in determining the required portal coupling across different reheating temperatures. For simplicity, we only highlight the muon and tau thresholds (as charm and bottom ones are very close to the tau threshold).

\begin{figure}[t!]
    \def\sepf{0.496}
    \centering
    \includegraphics[width=\sepf\columnwidth]{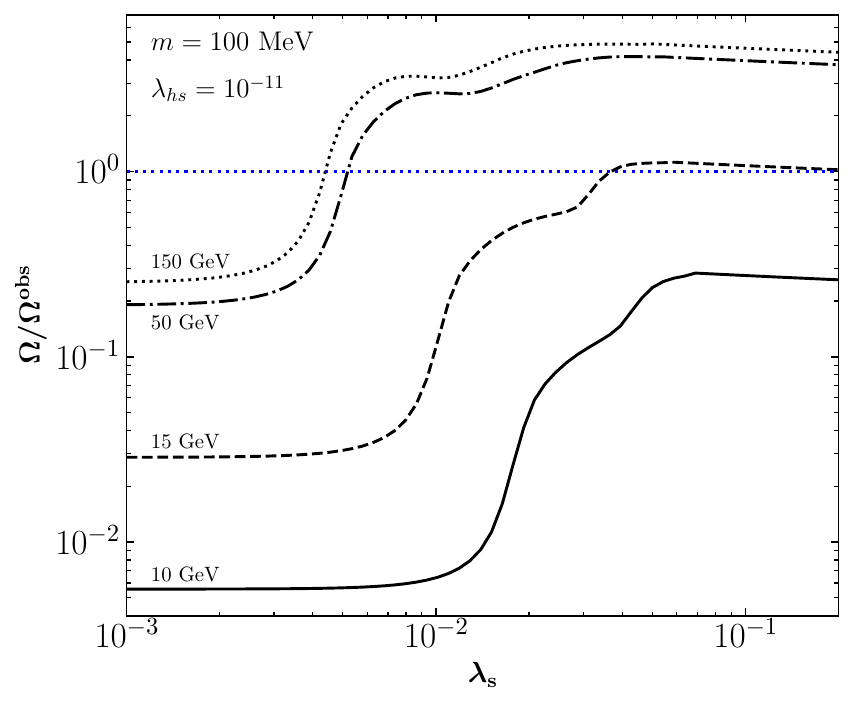}
    \includegraphics[width=\sepf\columnwidth]{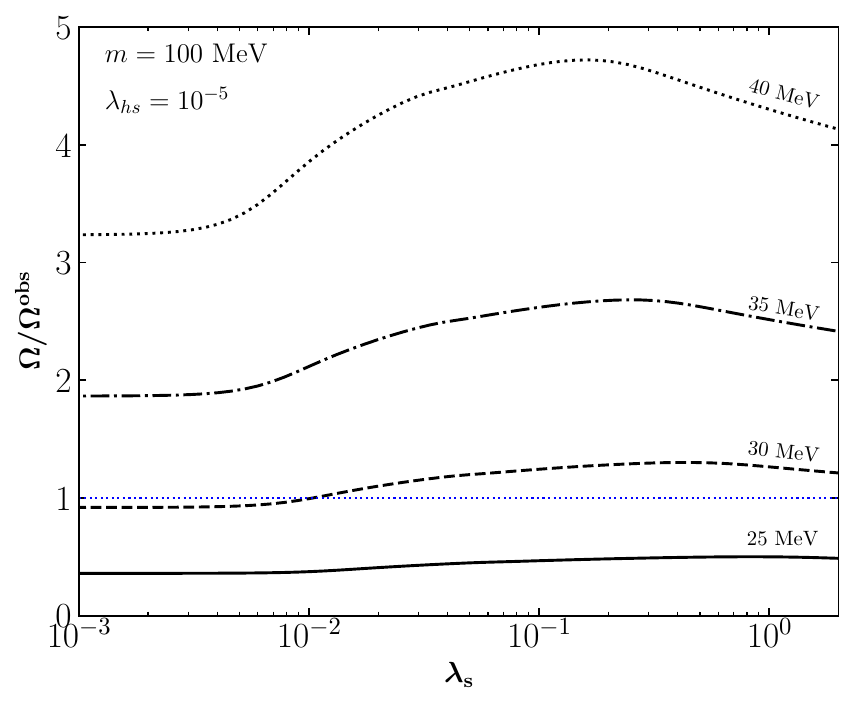}
    \caption{Final relic abundance as a function of $\ls$ for different reheating temperatures.} 
    \label{fig:ls_vs_relic}
\end{figure}
In Fig.~\ref{fig:ls_vs_relic}, we illustrate the ratio of the relic abundance obtained in our framework to the observed relic abundance as a function of $\ls$ for $m = 100$~MeV, considering two benchmark values of the portal coupling, and display contour lines for different $\Trh$ values. In the left panel, where $\lhs = 10^{-11}$, the small portal coupling requires large $\Trh$ values to approach the observed relic abundance. Furthermore, this suppressed portal interaction allows $\ls$ to significantly influence the abundance of relics at large values, since it enhances the production of DM through $2 \to 3$ self-interactions. This effect can increase the abundance by more than an order of magnitude. Once $\ls$ is large enough for self-thermalization to occur almost instantaneously, we see a plateau in the contours. In contrast, in the right panel with $\lhs = 10^{-5}$, the larger portal coupling leads to much smaller $\Trh$ values to satisfy the observed relic abundance. Moreover, the impact of $\ls$ is considerably diminished, as the enhanced portal interaction ensures that DM is produced more efficiently, rapidly bringing number-changing processes into equilibrium as $\ls$ increases. As a result, the production from $2 \to 3$ self-interactions becomes negligible. In fact, as $\ls$ approaches unity, the DM sector enters a phase where $3 \rightarrow 2$ interactions become increasingly efficient due to the already large DM population, leading to a decrease in the relic abundance.

\subsection{Prospects of light DM at future colliders}
\begin{figure}[t!]
    \def\sepf{0.496}
    \centering
    \includegraphics[width=\sepf\columnwidth]{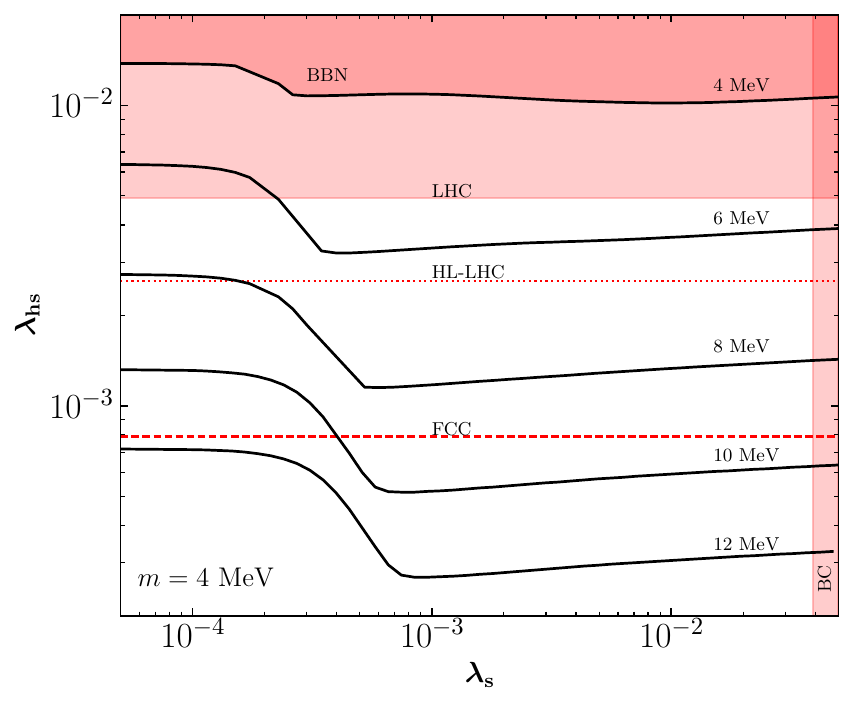}
    \caption{Values of $\Trh$ required to fit the whole DM abundance, in the plane $[\ls,\,\lhs]$, for $m = 4$~MeV.} 
    \label{fig:ls_vs_lhs}
\end{figure}
In Fig.~\ref{fig:ls_vs_lhs}, we present the detection prospects of a light DM candidate with mass $m = 4$~MeV in the $[\ls,\, \lhs]$ plane, with contours corresponding to different values of the reheating temperature $\Trh$. The red-shaded region above the $\Trh$ contour of 4 MeV is excluded by BBN, while the large red horizontal band is ruled out by existing LHC constraints. Limits from the Bullet cluster exclude the vertical red region on the right, arising from bounds on DM self-interactions. Future sensitivities of the HL-LHC and FCC, as previously indicated, are indicated by dotted and dashed red lines, respectively.

As an illustrative example, consider the contour for $\Trh = 6$~MeV. Although this scenario is excluded for negligible self-coupling $\ls$, increasing $\ls$ reduces the required portal coupling $\lhs$, bringing the model into a region that could be accessible in future collider experiments. This demonstrates how DM self-interactions with low $\Trh$ can open up viable regions in parameter space for a light DM that would otherwise remain excluded.

\section{Conclusions} \label{sec:concl}
In this work, we present the first comprehensive analysis of freeze-in dark matter (DM), incorporating the combined effects of a non-instantaneous reheating phase and dark-sector self-interactions. In particular, we studied the production of $\mathbb{Z}_3$-symmetric scalar DM with $3 \leftrightarrow 2$ number-changing {\it cannibal} processes.

Our study is based on the numerical solution of coupled Boltzmann equations governing the evolution of the number density and temperature of the DM sector, in conjunction with the thermal history of the SM bath. This approach reveals a rich and complex phenomenology that departs significantly from simplified scenarios assuming instantaneous reheating or negligible self-interactions. Through detailed benchmark analyses, we demonstrate that the DM thermal history is intricately influenced by the interplay between the reheating temperature ($\Trh$), the portal coupling to the Higgs ($\lhs$), and the DM self-coupling ($\ls$). 

For high reheating temperatures (e.g. $\Trh \simeq 1$~GeV), DM production is initially dominated by Higgs decays. Subsequent self-interactions within the dark sector can drive it into chemical equilibrium, initiating a characteristic cannibal phase where the dark sector temperature $T'$ exceeds the SM temperature $T$ once the DM becomes non-relativistic. In this regime, the freeze-out of self-interactions typically occurs after the completion of reheating.

In contrast, for low reheating temperatures (e.g. $\Trh \simeq 10$~MeV), significant entropy dilution requires much larger values of $\lhs$ to match the observed relic abundance. In this case, DM production from SM annihilations becomes dominant, and freeze-out of self-interactions can occur during the reheating epoch itself. The evolution of $T'$ in this regime is more complex, being shaped by continuous injection of energy from the SM bath and threshold effects such as hadronization.

Our parameter scans across the DM mass $m$, portal coupling $\lhs$, self-coupling $\ls$, and reheating temperature $\Trh$ lead to several key findings. The portal coupling required to yield the observed relic abundance is strongly dependent on $\Trh$; lower reheating temperatures demand significantly higher $\lhs$ to compensate for entropy dilution. Self-interactions play a crucial role, particularly for light DM. The 2-to-3 processes can significantly boost the DM abundance, enabling successful freeze-in with lower values of $\lhs$. The impact of $\ls$ is most pronounced for small $\lhs$ and large $\Trh$, where self-interactions can enhance the yield by over an order of magnitude. For larger $\lhs$, portal interactions become efficient enough to produce enough DM to rapidly establish equilibrium for number-changing processes, thereby reducing the sensitivity of the final yield to $\ls$. In such cases, strong 3-to-2 interactions can even deplete the DM density if overproduction occurs.

The viable parameter space is constrained by several experimental and observational bounds, including self-interaction limits from the Bullet cluster, Higgs invisible decay bounds from the LHC, lower bound on $\Trh$ from BBN, and direct detection experiments. We find that LHC constraints are already competitive with those from BBN and direct detection, and future projections from HL-LHC and FCC are expected to become the leading probes in much of the parameter space. This represents a significant improvement over conventional freeze-in models with large $\Trh$, where the requirement of extremely small portal couplings, typically $\lhs \sim 10^{-11}$, renders the dark sector practically invisible to current and next-generation experiments.

In particular, we identify scenarios in which the interplay between DM self-interactions and low reheating temperatures opens up previously excluded regions of parameter space, particularly for light DM with $m \sim \mathcal{O}(\text{MeV})$. In the absence of self-interactions, achieving the correct relic abundance in these scenarios would require large portal couplings, which are already excluded by collider constraints. However, a sufficiently large self-coupling $\ls$ can enhance DM production through number-changing processes, thereby allowing smaller portal couplings $\lhs$ to remain experimentally viable. These newly accessible regions offer promising prospects for detection in future collider experiments such as the HL-LHC or FCC. Importantly, they also provide an indirect window into the physics of low-temperature, non-instantaneous reheating, highlighting the broader cosmological significance of such scenarios.

In summary, our analysis underscores the necessity of accounting for both the detailed reheating dynamics and dark-sector self-interactions in evaluating freeze-in DM scenarios. The interplay between these factors can significantly alter the predicted relic abundance and reshape the viable parameter space, opening new directions for theoretical investigation and experimental searches for feebly-interacting DM.

\acknowledgments
NB received funding from the grants PID2023-151418NB-I00 funded by MCIU/AEI/10.13039 /501100011033/ FEDER and PID2022-139841NB-I00 of MICIU/AEI/10.13039/501100011033 and FEDER, UE. EC and AH are supported by the National Science Centre (Poland) under the research Grant No. 2021/42/E/ST2/00009.

\appendix
\section{Dark matter scattering with fermions} \label{ap:A}
The squared matrix element for the elastic scattering of DM with a SM fermion is 
\begin{equation}
    \vert\mathcal{M}_{Sf\to Sf}\vert^2 =N_c\, \lhs^2\, m_f^2\, \frac{4 m_f^2 - t}{(t - m_h^2)^2}\,. 
\end{equation}
For the temperatures of interest $T' \leq T < m_h$, and given the mass hierarchies $m\ll m_h$ and $m_f\ll m_h$, the transfer momentum is suppressed with respect to the Higgs mass squared, $m_h^2$. Therefore, $(t - m_h^2)^2 \simeq m_h^4$. Following Ref.~\cite{Cervantes:2024ipg}, we approximate the collision operator in the relativistic limit, while retaining the $e^{-m/T'}e^{-m_f/T}$ factors to account for the Boltzmann suppression at low temperatures. We notice that $4 m_f^2 - t =2 m_f^2 + 2E_2\, E_4 - 2\vert \vec p_2\vert\, \vert \vec p_4\vert\, \cos\alpha$, where $\alpha$ is the angle between $\vec p_2$ and $\vec p_4$, and we have labeled the momenta as $S_1 f_2\to S_3 f_4$. The constant part of the numerator results in
\begin{equation}\label{eq:ap1}
    C_2^{S f\leftrightarrow S f} \supset \frac{N}{N_\text{eq}}\frac{\lhs^2\, m_f^4\,N_c}{64 \pi^5\, m_h^4}\, T'^2\, T^2 \left(T - T'\right) e^{-m/T'} e^{-m_f/T}.
\end{equation}
We now evaluate the term with angular dependence,
\begin{align}
    C_2^{Sf\leftrightarrow Sf} &\supset \frac{2\, N_c\, \lhs^2\, m_f^2}{m_h^4} \nonumber\\
    &\times \int d\Pi_1\cdots d\Pi_4\, \frac{\vec {p_1}^2}{E_1} \left( E_2\, E_4 - \vert \vec p_2\vert\, \vert \vec p_4\vert\, \cos\alpha\right) \delta^{(4)}\left(\Sigma_f\, p_f - \Sigma_i\, p_i\right) \left(f_3\, f_4 - f_1\, f_2\right)\,,
\end{align}
and we proceed with the integrals over $p_1$ and $p_2$:
\begin{align}
    &\int d\Pi_1d\Pi_2\, \frac{\vec {p_1}^2}{E_1} \left( E_2\, E_4 - \vert \vec p_2\vert\, \vert \vec p_4\vert\, \cos\alpha\right) \delta^{(4)}\left(\Sigma_f\, p_f - \Sigma_i\, p_i\right) \nonumber\\
    &\qquad = \int d\cos\alpha\, \frac{1}{8\pi}\, \frac{\vert\vec p_1\vert}{\sqrt{s}} \left(\frac{ \vec {p_1}^2}{E_1}\right)^\text{cm} \left( E_1\, E_4 - \vert \vec p_1\vert\, \vert \vec p_4\vert\, \cos\alpha\right) \simeq \frac{(E_3+E_4)\, s}{16\pi}\,.
\end{align}
Here we have boosted the integrand to the CM frame and carried out the integration over $\vec p_1$ and $\vec p_2$, reducing the expression to an integral over $\cos\alpha$. In the relativistic limit, we approximate $E_i\simeq \vert \vec p_i\vert$. Note that the non-invariant term $\vec {p_1}^2/E_1$ transforms in the CM frame as
\begin{equation}
    \left(\frac{ \vec {p_1}^2}{E_1}\right)^\text{cm} = E_1\, z + \vert\vec p_1\vert\, \cos\alpha\, \sqrt{z^2 - 1} - \frac{m^2}{E_1\, z + \vert \vec p_1\vert\, \cos\alpha\, \sqrt{z^2 - 1}}\,,
\end{equation}
%
%
and we used $(\vec p_1/E_1)^\text{cm}\simeq E_1 (z+\sqrt{z^2-1}\cos\alpha)$ with the limit $\sqrt{z^2-1}\ll 3z$. Additionally, the remaining phase-space elements can be written as
\begin{equation}
    d\Pi_3\, d\Pi_4 = \frac{4\pi\, \vert\vec p_3\vert\, E_3\, dE_3}{2 (2\pi)^3\, E_3}\, \frac{4 \pi\, \vert\vec p_4\vert\, E_4\, dE_4}{2 (2\pi)^3\, E_4}\, \frac12\, d\cos\theta = \frac{1}{64 \pi^4}\, dE_3\, dE_4\, ds\,,
\end{equation}
where $\theta$ is the angle between $\vec p_3$ and $\vec p_4$ that enters the Mandelstam variable $s$ as
\begin{equation}
    s = m^2 + m_f^2 + 2\, E_3\, E_4 - 2\, \vert \vec p_3\vert\, \vert\vec p_4\vert\, \cos\theta\,.
\end{equation}
Evaluating the remaining integrals, we obtain the dominant contribution
\begin{align}
    \int dE_3 dE_4\, e^{-E_3/T'} e^{-E_4/T}\int_0^{4E_3 E_4} ds \frac{(E_3+E_4)s}{32\pi} = \frac{3}{2\pi} T'^3\, T^3\, (T+T')\,.
\end{align}
For the back reaction term (omitting overall numerical factors), we find:
\begin{align}
     \int dE_3 dE_4\, E_3e^{-E_3/T'} e^{-E_4/T}\int_0^{4E_3 E_4} ds \frac{s}{32\pi} = \frac{3}{\pi} T'^4\, T^3\,.
\end{align}
Reintroducing all prefactors, combining both contributions, we arrive at
\begin{equation}\label{eq:ap2}
    C_2^{S f\leftrightarrow S f} \supset \frac{N}{N_\text{eq}}\frac{3 \lhs^2\, m_f^2\,N_c}{64 \pi^5\, m_h^4}\, T'^3\, T^3 \left(T - T'\right) e^{-m/T'} e^{-m_f/T}.
\end{equation}
Finally, the sum of the contributions from Eqs.~\eqref{eq:ap1} and~\eqref{eq:ap2} gives
\begin{equation}
    C_2^{S f\leftrightarrow Sf} \simeq \frac{N}{N_\text{eq}}\, \frac{\lhs^2\, m_f^2\, N_c}{64 \pi^5\, m_h^4}\, T'^2\, T^2 \left(T - T'\right) \left(m_f^2 + 3\, T\, T'\right) e^{-\frac{m}{T'}}\, e^{-\frac{m_f}{T}},
\end{equation}
which is exactly the result in Eq.~\eqref{eq:C2_el_scatter_ferm}. 

\bibliographystyle{JHEP}
\bibliography{biblio}
\end{document}